\documentclass[12pt]{article}

\usepackage{graphicx}
\usepackage{amssymb}
\usepackage{amsthm}
\usepackage{booktabs}
\usepackage{rotating}
\usepackage{multirow}
\usepackage{eurosym}
\usepackage{amsfonts}
\usepackage{amsmath}
\usepackage{umoline}
\usepackage{caption}
\usepackage{color}
\usepackage{url}
\usepackage{authblk}
\usepackage[english]{babel}

\usepackage{pbox}
\usepackage{footnote}
\usepackage{multirow}
\usepackage{changepage}
\usepackage{tablefootnote}
\usepackage{ccicons}

\usepackage[margin = 2cm]{geometry}

\usepackage{hyperref}

\usepackage[natbibapa]{apacite}
\usepackage[symbol]{footmisc}
 
\usepackage{tikz} 
\usepackage{mwe} 
\usepackage{subfigure}
\usepackage{xcolor}
\usepackage[utf8]{inputenc}
\usepackage{import}
\usepackage{float} 
\usepackage{comment}
\usepackage{array}
\newcolumntype{P}[1]{>{\centering\arraybackslash}p{#1}}
\newcolumntype{M}[1]{>{\centering\arraybackslash}m{#1}}
\usepackage{tabularx}
\usepackage{dirtytalk}

\begin{document}

\title{On the dynamics of credit history and social interaction features, and their impact on creditworthiness assessment performance}
\footnotetext[1] 
{\scriptsize NOTICE: Changes resulting from the publishing process, such as peer review, editing, corrections, structural formatting, and other quality control mechanisms may not be reflected in this version of the document. This work is made available under a \href{https://creativecommons.org/licenses/by-nc-nd/2.0/}{Creative Commons BY-NC-ND license}. \ccbyncnd}
%\footnotetext[1]{E-mail addresses: %\href{mailto:rimunoz@uchile.cl}{rimunoz@uchile.cl} (Ricardo Muñoz-Cancino), \href{mailto:cbravoro@uwo.ca}{cbravoro@uwo.ca} (Cristi\'{a}n Bravo), \href{mailto:srios@dii.uchile.cl }{srios@dii.uchile.cl } (Sebasti\'{a}n A. R\'{i}os), \href{mailto:manuel.grana@ehu.es}{manuel.grana@ehu.es} (Manuel Graña)}

\author[1]{Ricardo Muñoz-Cancino}
\author[2]{Cristi\'{a}n Bravo}
\author[3]{Sebasti\'{a}n A. R\'{i}os}
\author[4]{Manuel Graña}

\affil[1,3]{Business Intelligence Research Center (CEINE), Industrial Engineering Department, University of Chile, Beauchef 851, Santiago 8370456, Chile}
\affil[2]{Department of Statistical and Actuarial Sciences, The University of Western Ontario,1151 Richmond Street, London, Ontario, N6A 3K7, Canada.}
\affil[4]{Computational Intelligence Group, University of Basque Country, 20018 San Sebasti\'{a}n, Spain.}

\date{}

\maketitle

\begin{abstract}
For more than a half-century, credit risk management has used credit scoring models in each of its well-defined stages to manage credit risk. Application scoring is used to decide whether to grant a credit or not, while behavioral scoring is used mainly for portfolio management and to take preventive actions in case of default signals. In both cases, network data has  recently been shown to be valuable to increase the predictive power of these models, especially when the borrower's historical data is scarce or not available. 
This study aims to understand the creditworthiness assessment performance dynamics and how it is influenced by the credit history, repayment behavior, and social network features.
To accomplish this, we introduced a machine learning classification framework to analyze 97.000 individuals and companies from the moment they obtained their first loan to 12 months afterward. Our novel and massive dataset allow us to characterize each borrower according to their credit behavior, and social and economic relationships.
Our research shows that borrowers' history increases performance at a decreasing rate during the first six months and then stabilizes. The most notable effect on perfomance of social networks features occurs at loan application; in personal scoring, this effect prevails a few months, while in business scoring adds value throughout the study period. These findings are of great value to improve credit risk management and optimize the use of traditional information and alternative data sources.

\end{abstract}


\section{Introduction}
Financial institutions operate in a complex and dynamic environment, where they are exposed to multiple risk sources, with credit risk being the most significant \citep{Apostolik2009}. 
Credit risk management is necessary and should be part of each lending decision; adequate risk management avoids financial losses and is a crucial element for the profitability and well-being of the financial institution and its borrowers \citep{Brown2014}.
One of the main objectives of credit risk management is to determine if the borrower will repay a loan and meet the agreed terms \citep{Basel2000}. For this, it is necessary to have policies, procedures, experience, and the expertise to extract knowledge from massive data sources \citep{Brown2014}. 
Researchers and practitioners have defined various types of credit scoring problems to manage credit risk, depending on the circumstances and background of each borrower \citep{Paleologo2010, Thomas2017}. In this research, we are interested in learning about application and behavioral scoring. Application Scoring supports the loan granting decision. Its objective is to assess the creditworthiness of new applicants; this is accomplished by combining the applicant's demographic information, loan repayment history, borrower historical data, and credit bureau data, along with data collected in the application form \citep{Anderson2022}. The credit risk management decision associated with application scoring is to grant loans to those borrowers who will be able to pay and avoid granting credits to those who will not.
Similarly, behavioral scoring models are used in credit risk management, but it is applied to existing customers \citep{Paleologo2010, Anderson2022}. In this way, all the loan payment behavior of the borrowers is available to develop an active portfolio management process. It enables lenders to take preventive actions with borrowers with high default likelihood, such as reducing the financial burden of those borrowers who have difficulties complying with the payment schedule and established obligations. 

Research on credit scoring is extensive but mainly focused on application scoring. Although some researchers describe behavioral scoring knowledge as limited and scarce \citep{Liu2001, Kennedy2013, Goh2019}, we are interested in delving into what we already understand about application scoring and behavioral scoring. First, both credit scoring problems combine borrower demographic data, historical information, and features obtained from multiple data sources. Repayment history emerges as one of the main creditworthiness predictors. The effect of this feature set is seen mainly in behavioral scoring; in application scoring, the payment behavior often is not available, or the applicant often does not have it \citep{munoz2021}. 
Second, better performance of credit scoring models leads to more accurate decision-making and allows more efficient and profitable credit risk management \citep{Verbraken2014, Djeundjie2021}. 
To increase the discriminatory power of these models, financial institutions have used alternative data, especially information from borrowers' relationships and interactions \citep{Ruiz2018, Oskarsdottir2019, Roa2021eswa, munoz2021}. This type of information adds value to both types of credit scoring. However, it is in application scoring where it achieves the most significant performance enhancement, especially in applicants whose repayment history is not available.

We know the effect of repayment behavior and social-interaction data on application and behavioral scoring problems based on the above. Concerning repayment behavior, this is gaining relevance as the relationship becomes entrenched; the more and more information on behavior is collected, the more accurately the borrower's creditworthiness can be predicted. In the case of social interaction data, the opposite occurs; at some point, the borrower's behavior and repayment history replace the knowledge provided by this alternative data source. 
Both relationships are engaging to study. However, to date, we do not know a study on the dynamics of this phenomenon. Research in credit scoring only knows what it occurs at the beginning (application scoring) and at some point during the loan payment schedule (behavioral scoring).

Consequently, this research endeavors to answer the following research questions:

\begin{enumerate}
    \item We know that borrowers' repayment history increases creditworthiness assessment performance. At what point does this information become meaningful? How long do we need to observe borrowers' repayment history? 
    \item We know that social interaction data adds higher value in application scoring when the behavioral information is scarce. How long is it convenient to use these sources of information? 
    \item What insights and value to credit risk management are obtained for studying this phenomenon?
\end{enumerate}

To this end, we have analyzed a massive multi-source credit dataset containing borrower information and social interaction data in the form of graphs. Then, we carried out an experiment where we observed individuals and companies from the moment they obtained their first loan their subsequent credit history repayment behavior for the next 24 months. The analysis of the results was carried out in considering credit history, repayment behavior, and alternative data and their impact on the creditworthiness assessment perfomance.

This work contributes to the growing research on credit scoring and the use of social network data. We challenge the current division of the credit risk management process through our analysis by investigating what happens between application scoring and behavioral scoring. Focusing the analysis on the borrower rather than the business process lets us discover how the credit scoring models' performance varies as the borrower begins to have a credit history. Additionally, we analyzed contribution repayment behavior features and social network features' contribution and how their value decreases in the presence of behavioral attributes.

Furthermore, our dataset is novel because it includes information for individuals and companies from the moment they obtain their first loan and their subsequent credit history repayment behavior, together with social network data to characterize them. In this way, it handles the low availability of data for behavioral models research stated by \citet{Kennedy2013} and \citet{Goh2019}. It allows us to carry out the first study, as far as we know, on credit assessment performance dynamics.

This paper is structured as follows. In the next section, the literature review of credit scoring in application and behavioral scoring models is presented. The proposed methodology is presented in Section \ref{sec:Methodology}. In Section \ref{sec:ExpDesign}, the experimental design and datasets are described.  Section \ref{sec:Results} is on results and their implications. The last section includes the conclusions, research findings, and future work.

\section{Background}\label{sec:relatedWork}

Credit scoring models enable and support credit risk management in financial institutions. For more than half a century, they have been part of the decisions throughout the credit risk management cycle  \citep{Thomas2017}. Decisions on whom to grant a loan to, portfolio management, preventive collection actions, and even pricing are not conceived today without the support of credit scoring models \citep{Bundi2016, Anderson2022}. Academics and practitioners have developed different credit scoring tools to address the different decisions at each credit risk management cycle stage. application scoring is used to decide whether to grant a loan to an applicant. In contrast, behavioral scoring allows to characterize those borrowers who have already been granted a loan, and it is used mainly for portfolio management. Finally, collection scoring allows optimizing the collection and recoveries policies and strategies \citep{Paleologo2010}.

The Application scoring models are used to decide whether to grant a loan or not. In this way, they are understood as the gateway to the lender institution and the financial system. Its correct implementation and usage allow the implementation of the risk appetite policies and define the applicant universe with whom to operate. Hence the importance of having models that allow to correctly quantify the borrower risk level and predict with high certainty whether the applicant will default or not. One of the most used approaches to enhance the creditworthiness assessment performance is to improve the modeling techniques, from the traditional logistic regression to other techniques such as support vector machines \citep{Huang2007}, bayesian models \citep{Kao2021}, genetic algorithms \citep{Kozeny2015}, ensemble classifiers \citep{Garcia2019, Radovic2021, Moscato2021}, and deep learning models \citep{West2000, Gunnarsson2021} and deep belief networks \citep{Luo2017, Gunnarsson2021}. While improving the performance is not restricted to an application scoring problem, most studies have been made on this particular problem. Another approach to enhance the creditworthiness assessment is to include alternative data sources. More and better data leads to better decisions, and in the case of application scoring, the researchers have analyzed the contribution of alternative data sources such as satellite and geospatial data \citep{Simumba2021},  psychometric data \citep{Shoham2004, Rabecca2018, Djeundjie2021}, mobile phone data and communication networks \citep{Oskarsdottir2018, Oskarsdottir2019}, network data \citep{Wei2015, Masyutin2015, Freedman2017, Cnudde2019, Giudici2020}, and written risk assessments \citep{Stevenson2020}. These studies have in common that most of the increase in creditworthiness assessment performance occurs when applicant traditional information is scarce or non-available.

Behavioral scoring is used mainly for portfolio risk management, understanding what happens after the credit has been granted. These models assess actual customers' creditworthiness and enable lenders to take preventive actions with borrowers with a high default likelihood. Unlike the application models, there are no extended research about behavioral scoring \citep{Liu2001, Goh2019}, \citet{Kennedy2013} suggest that the reason for scarce research on behavioral scoring is the large volume of data required and the difficulty of accessing the data. 
However, the research lines to increase the performance of these models are the same as in application scoring. \citet{Septian2020} investigated the value of social network data to predict bankruptcy, and \citet{Letizia2019} included a corporate payments network to assess an internal rating. Our previous research \citep{munoz2021} shows that social network data generate a much more significant performance enhancement in application scoring than behavioral scoring, considering the same population and features. Moreover, this result is consistent when the credit scoring model is applied to both individuals and companies, personal scoring and business scoring, respectively. 

Among other studies that address the behavioral scoring problem, the work of \citet{Hsieh2004} developed a behavioral scoring to manage credit card customers through an RFM-based segmentation model and then defined marketing strategies for each group using association rules. \citet{Biron2013} studied what happens when infringing the logistic regression independence assumptions in behavioral scoring. \citet{Kao2021} concluded that increasing the APR (annual percentage rate)  significantly increases the probability of default using the credit cardholder database from Taiwan.

The Behavioral scoring models include additional information: repayment behavior, and credit history \citep{Thomas2000}; this data is not necessarily available in application scoring. The repayment behavior and banking data preceding the observation point is defined as the performance period \citep{Thomas2000}. The behavior during this period is added as features - for instance, number of missed payments and average balance.
There is still no consensus on the performance period length. \citet{Thomas2000, Liu2001} gives 12 months as an example and \citet{Djeundjie2021} stated \say{Behavioural scoring models are applied to accounts that have been open for a sufficient period} (p.2), but without details of the sufficient period. \cite{Kennedy2013} analyzed the performance period length comparing windows of 6-months, 12-months, and 18-months. Concluding the best performance is achieved using the 12-months performance windows but limited to shorter outcomes windows; in longer outcome windows, it is harder to find optimum performance windows. Therefore, the performance window selection only affects the short-term creditworthiness assessment.

Despite all of this, we do not fully understand how long the performance period should be and how the discrimination power varies as more knowledge on repayment behavior becomes available. Additionally, the role and contribution of network data in the shift between application and behavioral scoring remain an open question.

\section{Methodology}\label{sec:Methodology}
We use an approach based on machine learning classification models to analyze the creditworthiness assessment performance's dynamics and how it is affected by credit history, repayment history, and social network features. 

We analyze the performance dynamics by studying multiple machine learning static models; this is achieved by creating twelve credit scoring datasets varying the number of months since the borrower obtained his first credit. The first dataset only includes borrowers in the first month after granting. The first dataset corresponds to a credit scoring application problem; the second dataset includes the same group of borrowers, but they are observed two months after granting. It continues until the last dataset contains the borrowers' information 12 months after granting. 
It is necessary to point out that although the models to be trained are independent, the training data is not. In this way, it is possible to obtain insights into performance dynamics through multiple static models.
The borrowers, individuals, and companies in the study remain invariant during the 12 months of analysis. As the months go by after the loan is granted, borrowers can be excluded from the training dataset due to default or full credit payment. Borrowers are described using the same feature set, regardless of the analysis time. However, these features reflect diverse behaviors as the borrower repays the loan or shows signs of credit deterioration.

This study uses gradient boosted trees because they have consistently shown state-of-the-art performance in different problems \citep{Friedman2001, Tianqi2016, munoz2021}. Additionally, in order to quantify the performance, we use the area under the receiver operating characteristic curve \citep[AUC]{Bradley1997}  and the Kolmogorov-Smirnov statistic \citep[KS]{Hodges1958} as performance measures.

The dataset used to train these models is built by applying the feature engineering process described in section \ref{sec:featEng}. To obtain insights based on the generalization capability of these models and avoid overfitted models, we split the dataset into two parts. The first dataset that contains 30\% of the original dataset is used for feature selection and hyper-parameter optimization. The remaining dataset is used to train the final models, including the features and parameters previously obtained. The results and conclusions are based on the average of a 10-fold cross-validation. The comparison between models is made using a t-test, according to the recommendations introduced in \citet[Ch.\ 12]{Flach2012}. 

The feature selection and hyper-parameter optimization start discarding those features with low or almost null predictive power; for this, we calculate the KS and AUC in a univariate way and discard those attributes with a $KS <= 0.01$ or an $AUC <= 0.53$ and then apply a method to drop out highly correlated features. This method begins by selecting the feature with the highest predictive power and then discards those whose absolute value of the correlation is greater than a parameter $\rho = 0.7$; this process is repeated until all the target features are evaluated. We use this method twice, first considering attributes that belong only to the feature sets defined in \ref{sec:featEng}; thus, we ensure a representative mix of attributes for each dimension analyzed. Then, we apply it again by considering all the remaining features. Finally, to find the best hyper-parameters, we apply an exhaustive search over specific parameters, the number of boosted trees to fit and their learning rate, and the minimum data needed in leaves.

Our methodology's second and last stage consists of using the remaining 70\% of the original dataset to train models using N-Fold Cross-Validation together with the previously obtained hyper-parameters and features.

\section{Experimental Setup}\label{sec:ExpDesign}
\subsection{Dataset}\label{sec:datadescr}

We have used data provided by a Latin American bank. The information was anonymized to guard customer confidentiality and not compromise any customer's identification and relationships; there is no possibility that this study can leak private data. 
The data includes 97,044 customers who obtained their first loan between January 2018 and December 2018 and contains their repayment behavior until December 2019. Our approach involves building a machine learning classification model. For this, each borrower is labeled as a defaulter or non-defaulter. A borrower to be considered a defaulter must be more than 90 days past due within the next 12 months from the moment it sampled; if during this outcome period the borrower is not more than 90 days past due, it is considered non-defaulter. Those borrowers with arrears of more than 90 days at the time of observing them are not considered in the analysis.

The data sources used to construct this dataset have already been used to develop credit scoring models \citep{munoz2021}. Application and behavioral scoring models were trained. In both cases, the borrower information and social interaction data proved to be a statistically significant contribution to increasing creditworthiness assessment performance. Additionally, in this study, we are interested in analyzing credit scoring models according to the type of borrower. We will refer to personal credit scoring when borrowers are individuals and business credit scoring when they are companies. This classification is complementary to the one previously defined. In this way, it is possible to assess individuals' creditworthiness (personal credit scoring) at the application or after that, i.e., application or behavioral scoring, and the same applies to companies. 

Table \ref{table:datasetdesc}  describes the available information; this is grouped into business credit scoring data and personal credit scoring data to distinguish between companies and individuals. It shows the borrower number in the first observation month and the features number corresponds to those provided by the financial institution and those built for this research. Each of the 12 datasets is characterized based on the same attributes and contains the same borrowers observed $i$ months ($i = 1, \ldots, 12$) after their loans were granted. 

\begin{table}[ht]
\footnotesize
\centering
\caption{Dataset description. Borrowers correspond to the total number of individuals and companies that are part of our analysis, which will be observed from the moment they obtain a loan until 12 months later.}
\label{table:datasetdesc}
\resizebox{0.5\textwidth}{!}{
\begin{tabular}{|p{4cm}|c c|}
\hline
{\centering Model} & {\centering Borrowers} & {\centering Features} \\ \hline
Business Credit Score & 20,835 & 585 \\ 
 Personal Credit Score & 76,209& 936\\ \cline{1-3}
\end{tabular}}
\end{table}

\begin{figure}[ht]
\centering
\resizebox{0.80\textwidth}{!}{%
\input{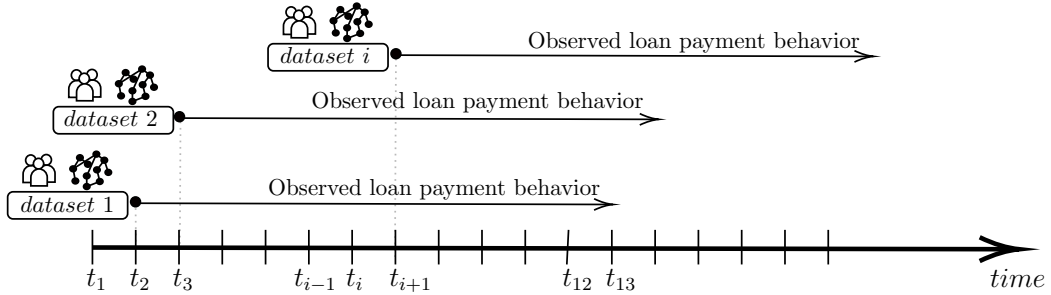}}
\caption{Dataset construction}
\label{fig:expSetup}
\end{figure}

An overview of the construction of the datasets can be seen in Figure  \ref{fig:expSetup}. The timeline shows how many months have passed since the credit was granted; in this way, $dataset_i$ contains all the borrowers after $i$ months having obtained credit. During the following 12 months from the observation month $i$, their payment behavior is analyzed to label the borrower as a defaulter or non-defaulter if they have exceeded 90 days in arrears during this observation period.

\begin{figure}[ht]
  \centering
  \noindent
  \resizebox{1\textwidth}{!}{
  \begin{tabular}{cc}
    \subfigure[Business Scoring]{\includegraphics[width=0.5\linewidth]{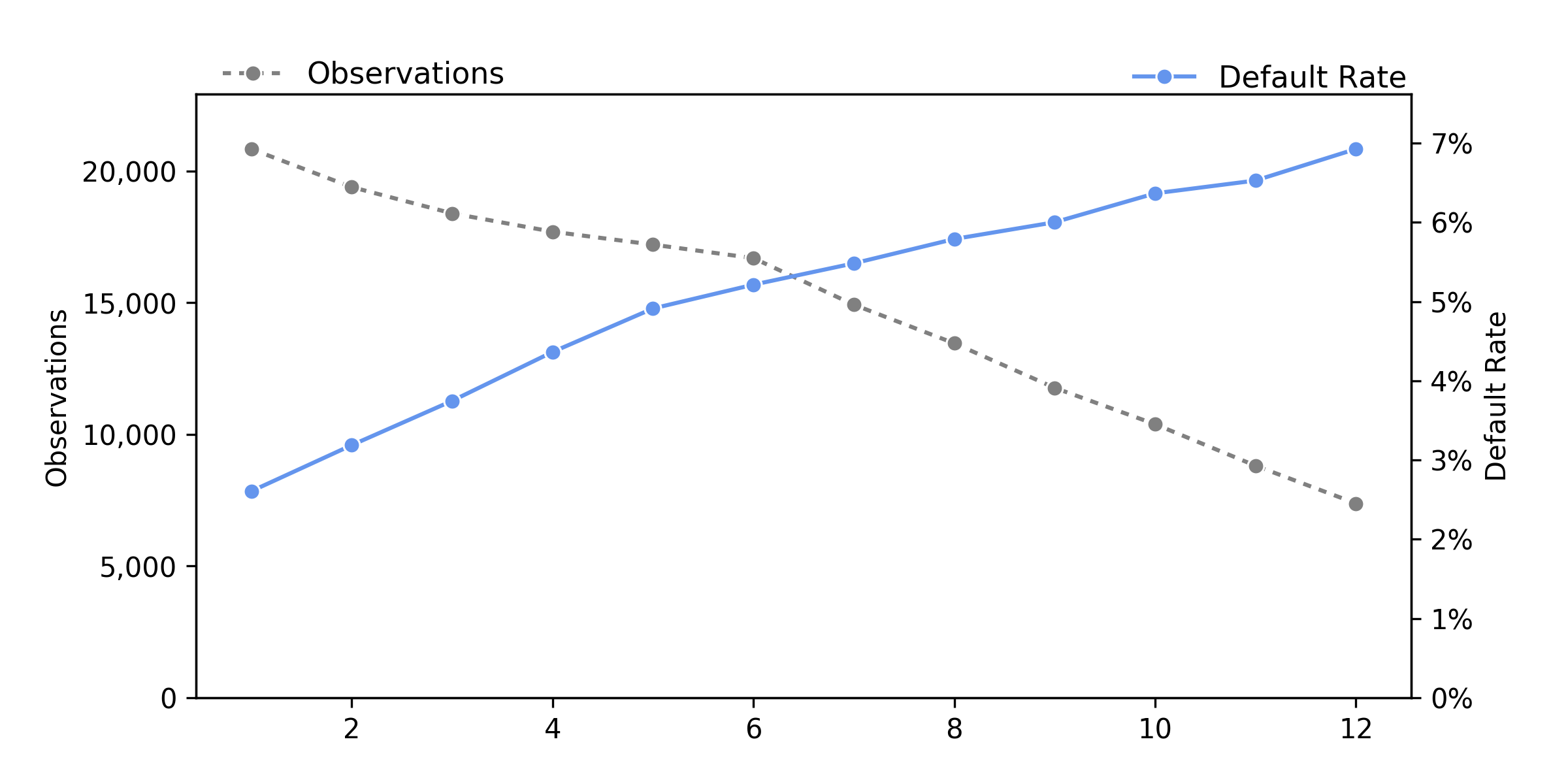}}&
    \subfigure[Personal Scoring ]{\includegraphics[width=0.5\linewidth]{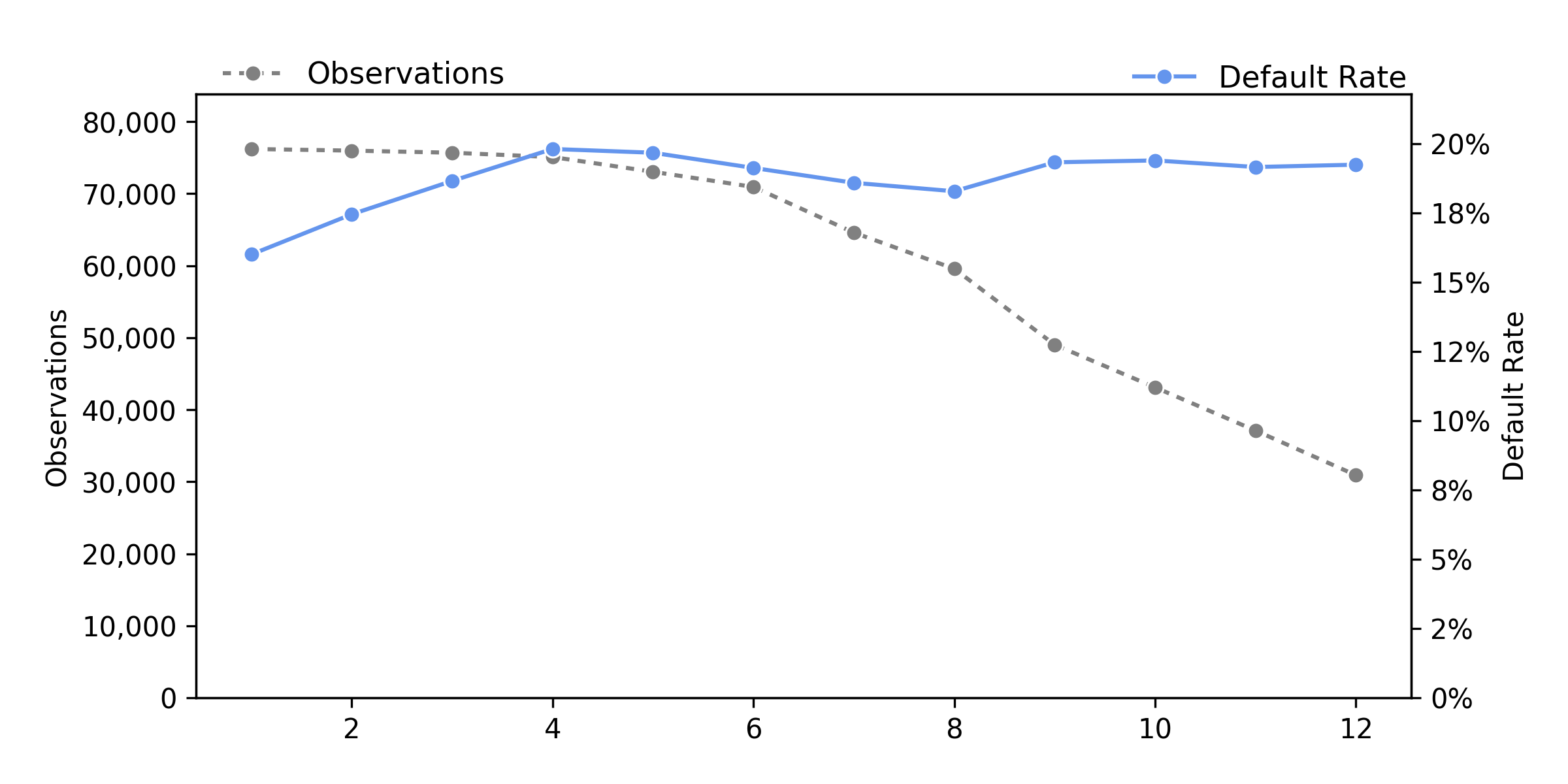}}
  \end{tabular}}
    \caption{Dataset Statistics. The X-axis displays the number of months elapsed since the loan granting. The left Y-axis shows the observation samples, and the right Y-axis the default rate.}
    \label{fig:datasetdesc}
\end{figure}

Figure \ref{fig:datasetdesc} shows the borrower's number and the default rate for each of the 12 datasets, grouped by business scoring and personal scoring.  Both charts start with borrowers at the moment of application, 20,835 and 76,209 borrowers for business and personal scoring, respectively. These borrowers are the ones that are part of the 12 dataset. The observations number decreases for two reasons, the default or the total payment of the loan. These datasets are mutually dependent since they contain information on the same borrowers but with diverse progress in repaying their loans. This aspect allows us to gain insights into performance dynamics from independently trained models.

\subsubsection{Traditional data sources: Borrower Data}\label{sec:nodedata}
To describe the borrowers, we have a massive dataset with the financial information of 7.65 million people and almost a quarter-million companies between January 2018 and March 2020. The financial information provided in this dataset corresponds to the monthly debt decomposition by type and by days past due grouped into buckets. This particular dataset contains all of the study subjects described in Table \ref{table:datasetdesc}.

\subsubsection{Alternative data sources: Social Interaction Data}\label{sec:networkdata}

This study will characterize the companies with network data information originating from their economic and social interactions. The network used for this purpose is composed of Transactional services, the Enterprise's ownership, and the Employment relationship. This network tries to build an ecosystem where companies, business owners, and employees interact. We will call this network \textbf{EOWNet}. On the other hand, people will be mainly characterized by combining marriages, parents, and children. We will call this network \textbf{FamilyNet}. The EownNet is also used in the personal scoring problem since many of these borrowers are part of the EOWNet. However, due to the partial coverage of this dataset, it is expected a limited added value as we observed in \citet{munoz2021}.
The \textbf{EOWNet} is a dynamic network since the interactions that constitute it changes monthly, while the \textbf{FamilyNet} is a static network, fixed at the beginning of the study period.

\subsection{Borrower and social network features}\label{sec:featEng}
The people and companies of this study will be characterized by features created through a feature engineering process. This process combines the node information, the repayment history \ref{sec:nodedata} and the network data \ref{sec:networkdata}. We will classify these features into the following subset: 

\begin{itemize}
    \item \textbf{Borrower's Financial Features:} Correspond to borrower features based on the information provided by the financial institution and allows us a characterization of the financial situation of each borrower. Contains the debt decomposition by type (consumer, commercial, and mortgage) and by delinquency situation, and the amount in revolving loans. We will call this feature set $X_{Fin}$. It exclusively represents the borrower's situation at the moment of observation. Additionally, we include another feature set with the borrower's historical information and their repayment history; we will call this feature set $X_{FinHist}$. This borrower's historical features include the mean and SD for each $X_{Fin}$ feature for the last 3 and 6 months. In this way, both $X_{Fin}$ and $X_{FinHist}$ feature set describes the borrower's financial situation. However,  $X_{Fin}$ describes the borrower's credit situation at the observation point, and $X_{FinHist}$ summarizes the historical borrower's financial situation for the last 3 and 6 months.  
    
    \item \textbf{Node Statistics:} This feature set considers each borrower as a node within the network. Therefore, these features correspond to nodes' statistics based on their position and characteristics within the network. For each node in the network, we calculate the Degree, Degree Centrality, number of Triads, PageRank Score, authority and hub score from hits algorithm and an indicator of whether the node is an articulation point.  We will call this feature set $X_{NodeStats}$, and it is part of the features that utilizing alternative data sources provides us.
    
    \item \textbf{Social Interaction Features:} We utilize the borrower's social interactions to characterize each borrower based on their neighborhood's financial information, i.e.,  the individuals and companies to which they are connected. Formally, we use the borrower ego network to characterize a borrower using social network data corresponding to all the nodes the borrower is connected to. We aggregate the egonet financial features in the $X_{SocInt}$ feature set, calculating the mean and SD for the nodes' features in the borrower's egonet \citet{Nargesian2017, Roa2021eswa}. As we did with borrowers' features, we aggregate historical social interaction features by calculating the mean and SD of the last 3 and 6 months. We will call this additional feature set $X_{SocIntHist}$.

\end{itemize}

\subsection{Experiments} \label{sec:RfeatSetup}
We define a series of experiments to analyze the performance dynamics effect of credit history, repayment history, and social network features. For this, we generate different sets of characteristics detailed in Table \ref{table:featuresGroup}. With each of these feature sets will train twelve models, one for each month that has elapsed since the loan granting.

\begin{table}[ht]
\footnotesize
\centering
\caption{Experiments Setup}
\label{table:featuresGroup}
\begin{tabular}{cl}
\toprule
   Experiment Id  &    Feature Group\\ \midrule
    E1 &  $X = \{ X_{Fin} \}$ \\
    E2 &  $X = \{ X_{Fin} + X_{FinHist}\}$ \\
    E3 & $ X = \{ X_{Fin} + X_{FinHist} + X_{NodeStats} + X_{SocInt} + X_{SocIntHist}\}$ \\ \bottomrule    
\end{tabular}
\end{table}

\section{Results and Discussion}\label{sec:Results}

This section presents the results obtained after applying our methodology to study creditworthiness assessment performance's dynamics. First, we present how the borrower's credit history affects the model's performance (Experiment \textbf{E1}). Then we show how this effect changes when the repayment features are incorporated into the analysis (Experiment \textbf{E2}). Finally, we study the effect on the model's performance incorporating the social interaction features, the borrower's credit history and repayment features (Experiment \textbf{E3}). 

A further relevant analysis is to understand how much the social interaction features influence the creditworthiness assessment compared to the borrower's features and how this impact varies over time. The results of this analysis are presented in \ref{res:featureImportance}.

\subsection{Experiment \textbf{E1}:  borrower credit history}

\begin{figure}[ht]
  \centering
  \noindent
%   \vspace{-0.5cm}
  \resizebox{0.85\textwidth}{!}{
  \begin{tabular}{cc}
    \subfigure[Business Scoring (KS)]{\includegraphics[width=0.5\linewidth]{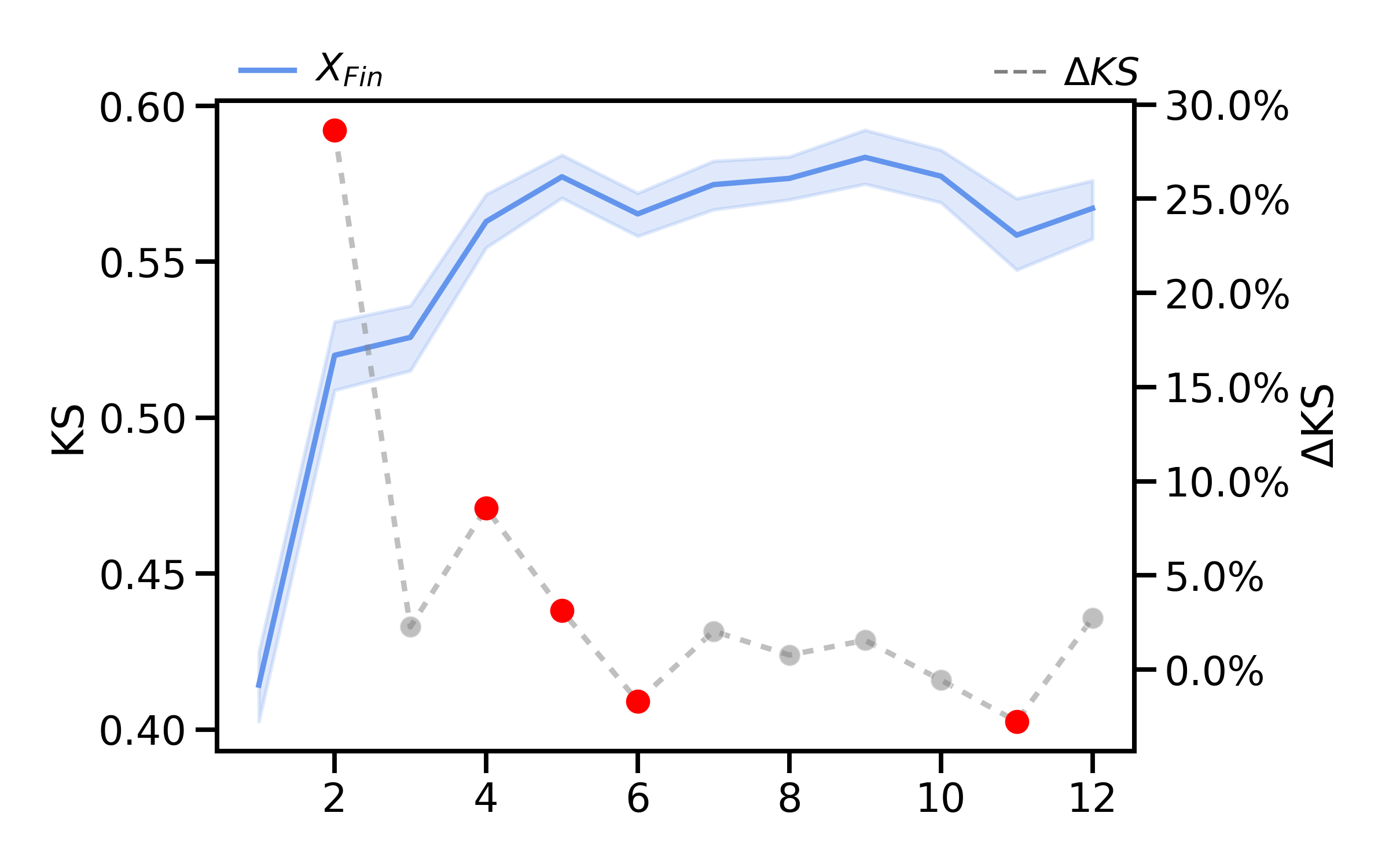}}&
    \subfigure[Business Scoring (AUC)]{\includegraphics[width=0.5\linewidth]{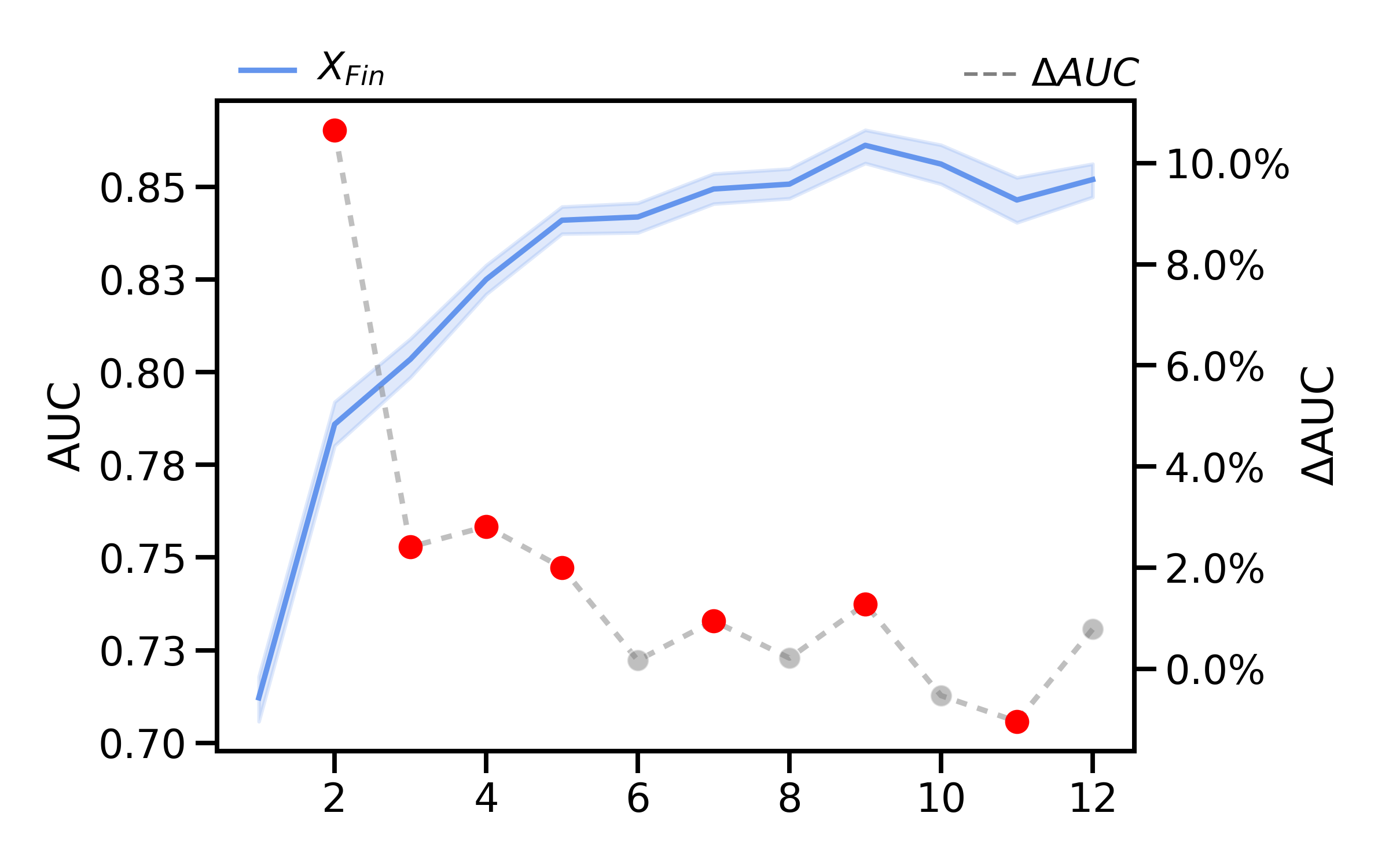}} \\
        \subfigure[Personal Scoring (KS)]{\includegraphics[width=0.5\linewidth]{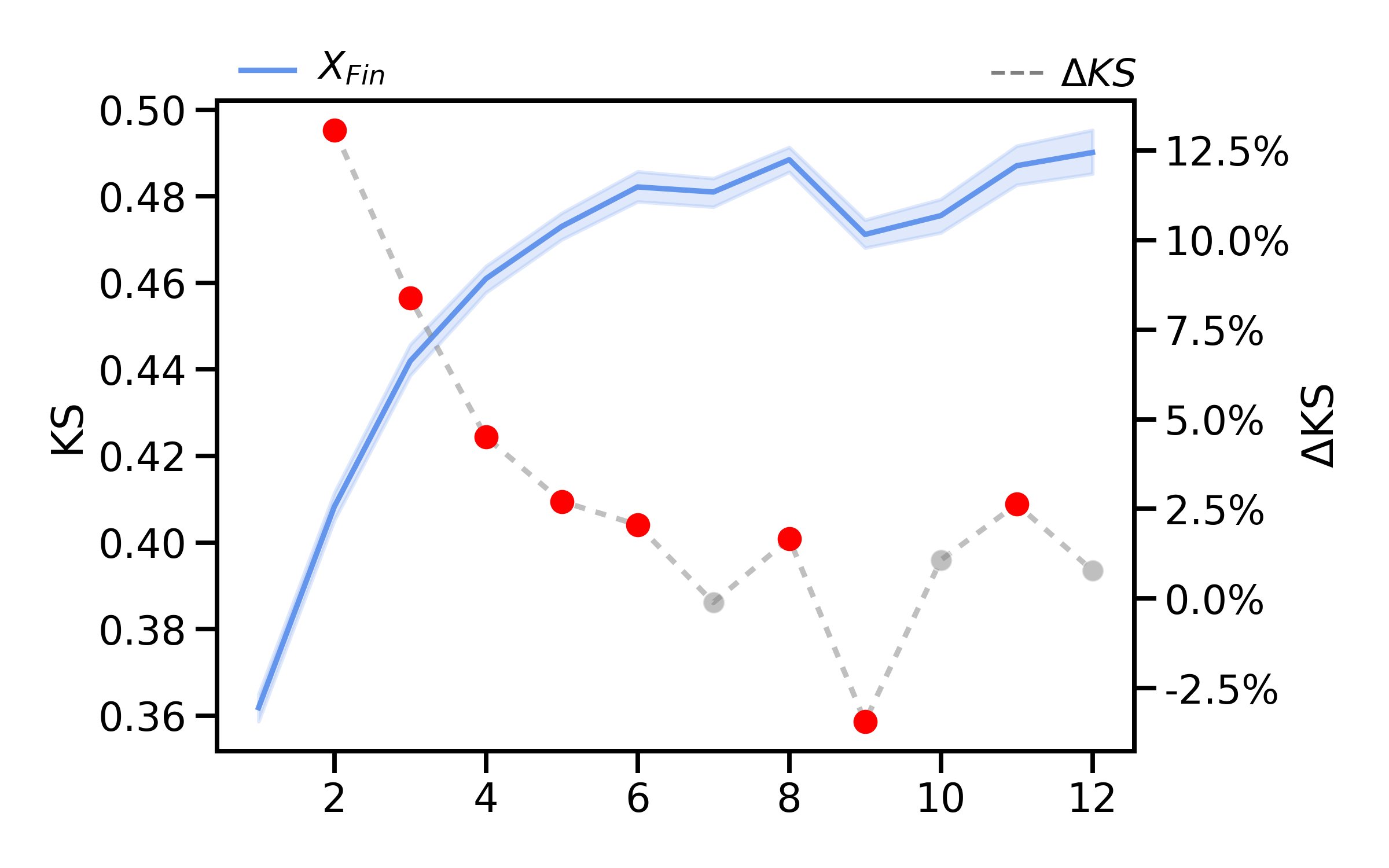}}&
    \subfigure[Personal Scoring (AUC)]{\includegraphics[width=0.5\linewidth]{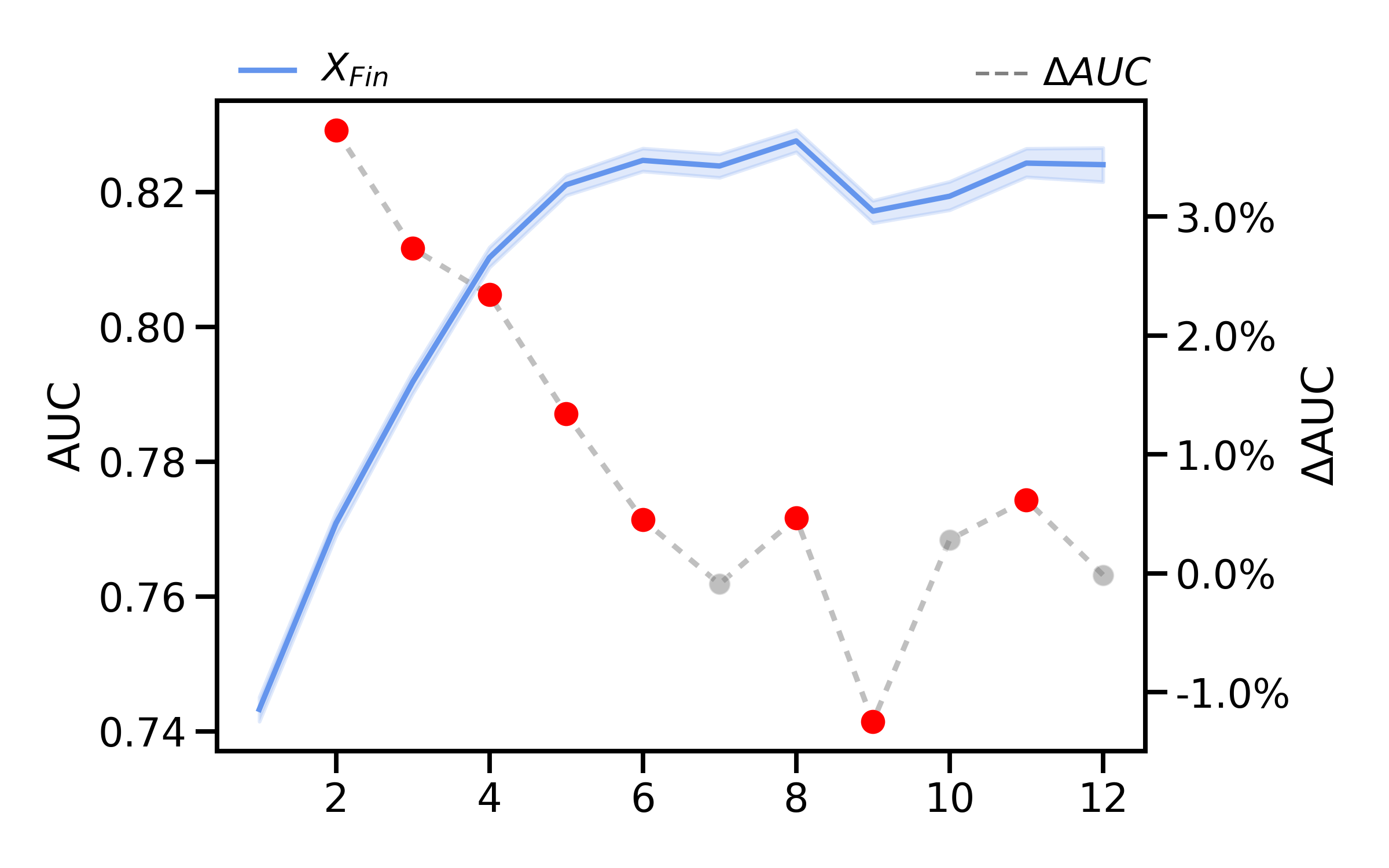}} 
  \end{tabular}}
     \caption{Results in terms of KS and AUC scores for the Business Scoring and Personal Scoring Problem. The X-axis displays the number of months elapsed since the loan granting. The blue line shows the creditworthiness assessment performance (left Y-axis) for the Experiment \textbf{E1}, using only $X_{Fin}$: Node Features. The dotted gray line (right Y-axis) shows the percentage increment between consecutive periods; when this increment is statistically significant, the dots are colored red. Otherwise, they are colored gray.}
    \label{fig:NodeHistory}
\end{figure}

The first goal is to understand how the borrower's credit history affects the creditworthiness assessment performance. We will do this by analyzing the behavior of the borrowers' financial features $X_{Fin}$ over time. The Figures \ref{fig:NodeHistory}(a) and \ref{fig:NodeHistory}(b) show this effect for the business scoring problem and the  Figures \ref{fig:NodeHistory}(c) and \ref{fig:NodeHistory}(d) for personal scoring. For each problem, performance is evaluated using the KS and AUC scores.

For each elapsed month since the loan granting, the initial feature set at the beginning of the training is the same. However, as after applying the methodology defined in Section \ref{sec:Methodology}, the final variables may vary between one period and another since we select those that better improve default assessment.
It is observed that the discriminatory power increases as the borrower credit history increases, and this increment is at a decreasing rate. Furthermore, the increase ceases to be consistently statistically significant after six months, so the gains in discriminatory power are relevant in the first months.

In business scoring, we observe a high increase in the second month, higher than 25\% in KS and 10\% in AUC; however, additional credit history produces relatively minor increases. Something different happens in personal scoring, where the increases, although lower, are consistent in the first six months.

These results confirm what academics and practitioners already know: the importance of borrower credit history in the creditworthiness assessment. The value of these results is that they reveal the discrimination power dynamics produced by the availability borrower history. The credit history produces increases in performance at a decreasing rate. After six months, the gains are marginal; this suggests that the transition from an application scoring problem to a behavioral scoring problem, in terms of discriminatory power, occurs in these six months. Our results challenge the imprecise definition of the performance window in behavioral scoring from a sufficient period \citep{Djeundjie2021} or the proposed 12 months \citep{Thomas2000, Liu2001, Kennedy2013} and reduce them to the six-month performance window observed in this study.

A smaller performance window allows reducing the volume of information necessary to research behavioral scoring models, which, as we have seen, is a limitation in this area \citep{Kennedy2013, Goh2019}. Furthermore, these results broaden the adequate target population to apply these models, going from borrowers with 12 months of credit history to needing only six months.
The third point favoring a six-month window performance is that it allows faster defaulter recovery. That is to say, the credit evaluation for borrowers who had negative events in the past and now present a good payment behavior, being this of great help in financial inclusion and access at lower interest rates. Finally, other benefits include more straightforward technological implementations, reducing storage costs, and generating behavior models that quickly capture the portfolio's trends and shifts.

\subsection{Experiment \textbf{E2}:  borrower credit history and repayment features}

Another advantage of the borrower's credit history is to build attributes that reflect the temporal evolution of its characteristics. For this, we create a set of features that summarize the credit information of the last three and the last six months. In the first period of analysis, these attributes do not add value since there is no previous history; however, these attributes make sense as the months' pass since loan granting.

\begin{figure}[ht]
  \centering
  \noindent
  \resizebox{0.85\textwidth}{!}{
  \begin{tabular}{cc}
    \subfigure[Business Scoring (KS)]{\includegraphics[width=0.5\linewidth]{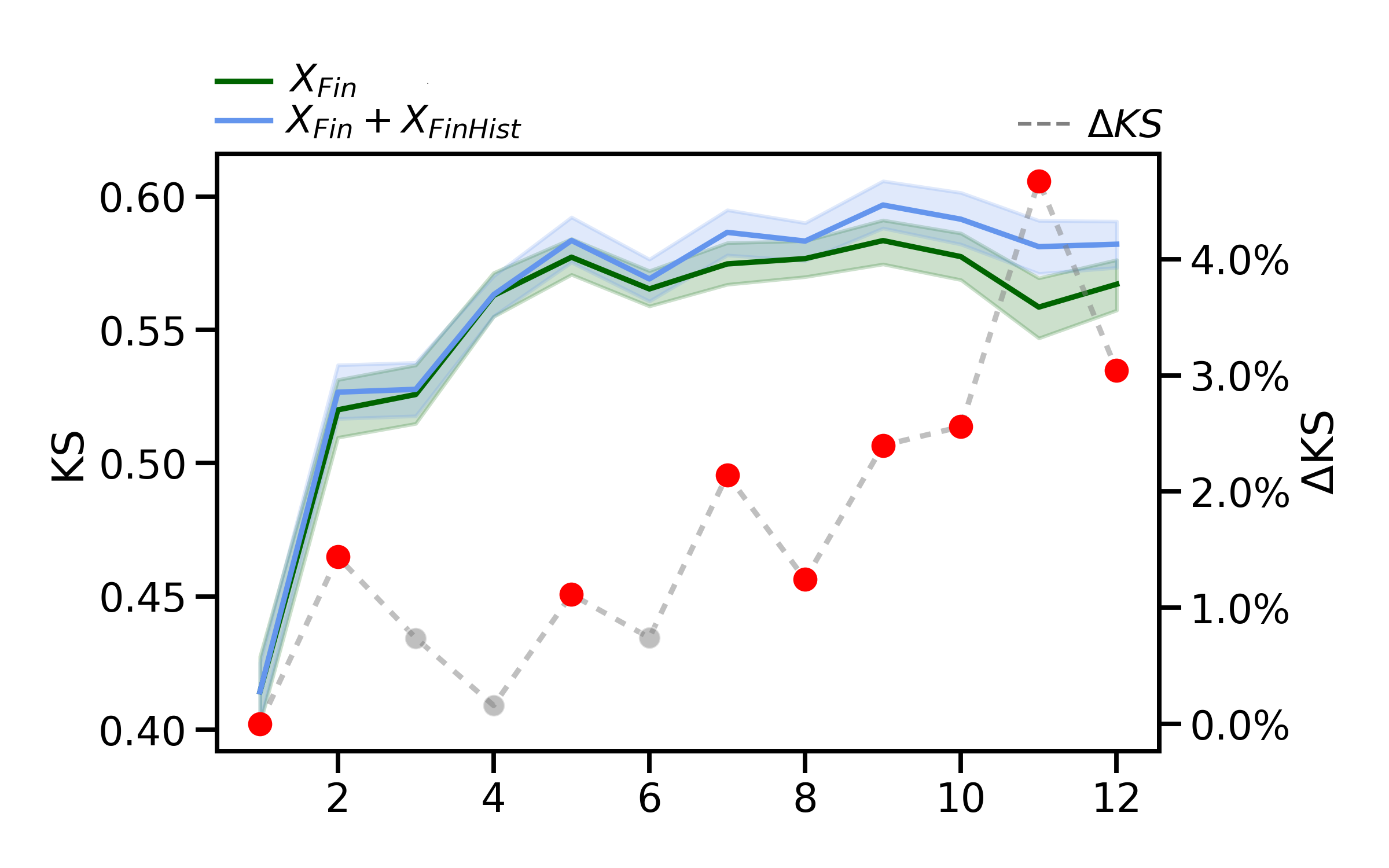}}&
    \subfigure[Business Scoring (AUC)]{\includegraphics[width=0.5\linewidth]{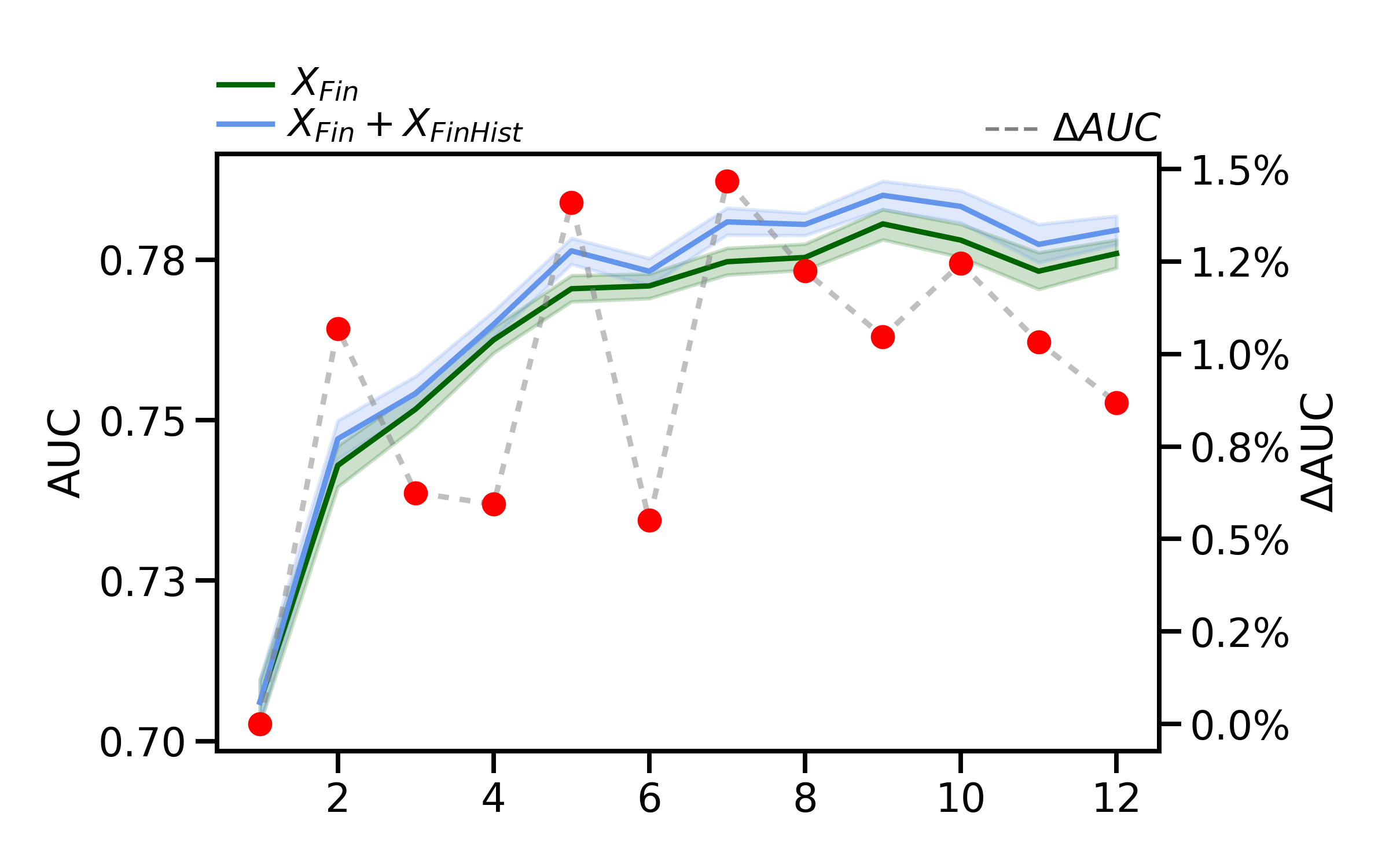}} \\
        \subfigure[Personal Scoring (KS)]{\includegraphics[width=0.5\linewidth]{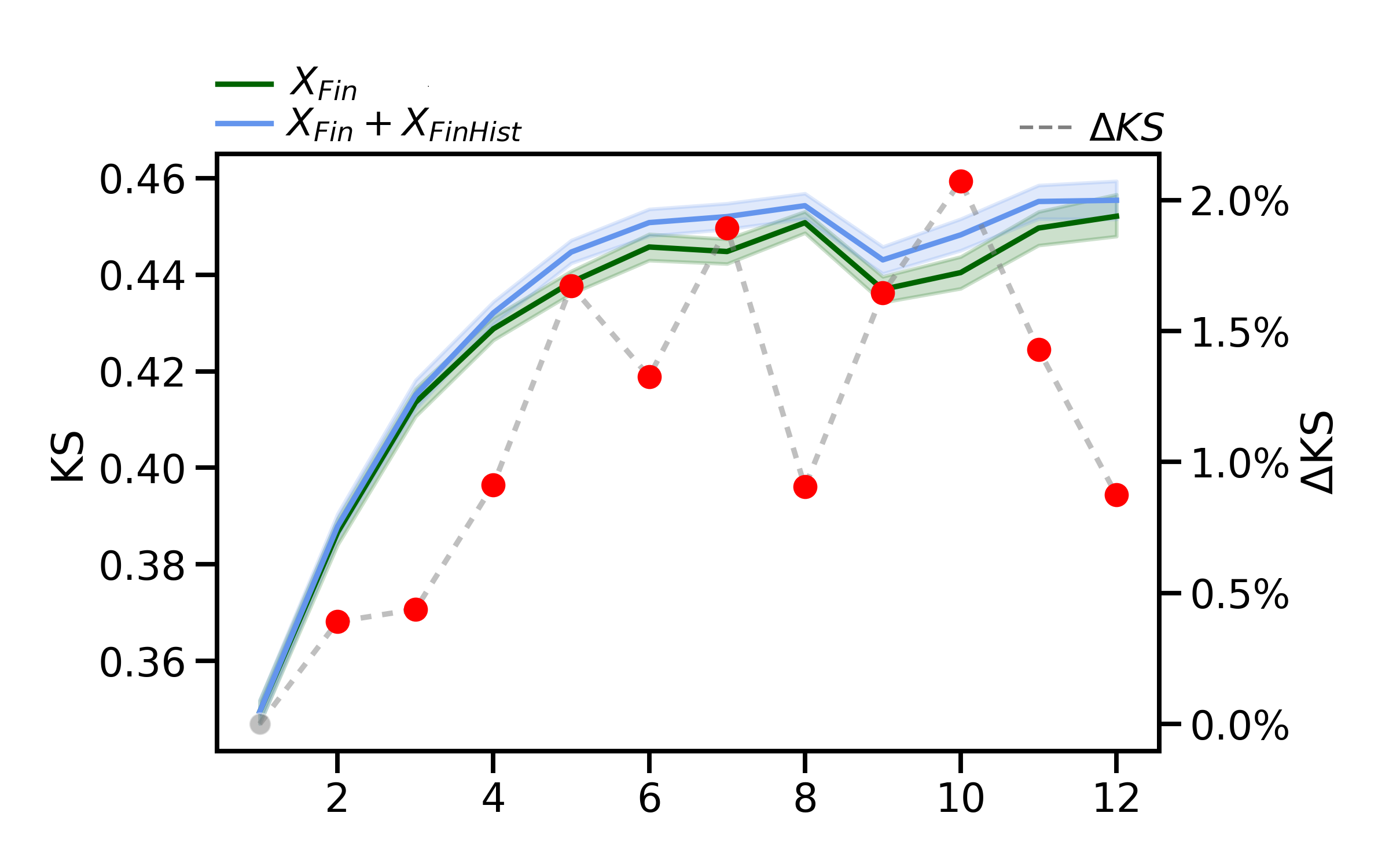}}&
    \subfigure[Personal Scoring (AUC)]{\includegraphics[width=0.5\linewidth]{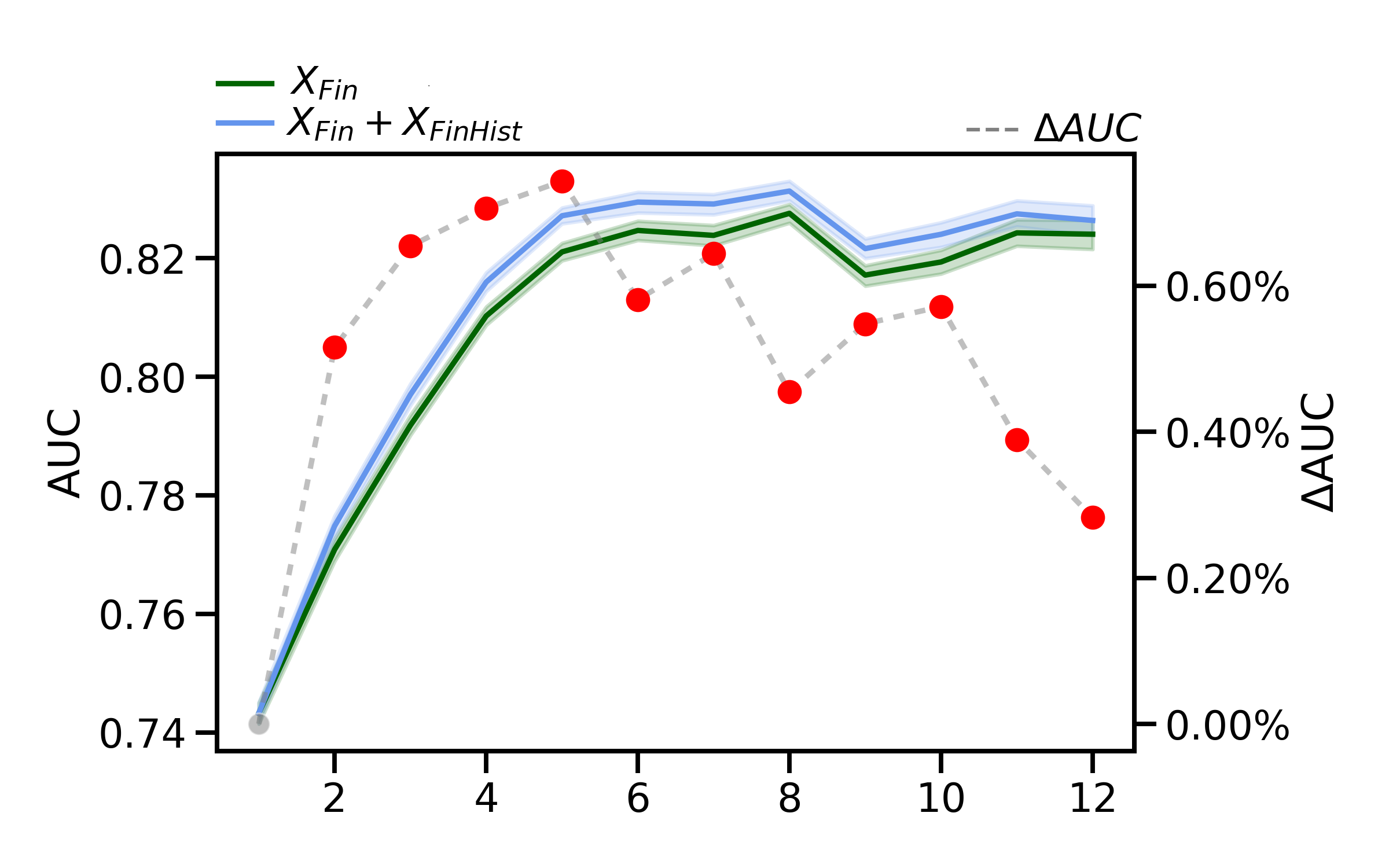}} 
  \end{tabular}}
   \caption{Results in terms of KS and AUC scores for the Business Scoring and Personal Scoring Problem. The X-axis displays the number of months elapsed since the loan granting. The blue line and the green line show the creditworthiness assessment performance (left Y-axis) for Experiment \textbf{E2} and Experiment \textbf{E1}. The dotted gray line (right Y-axis) shows the percentage increment between  \textbf{E2} and \textbf{E1}; when this increment is statistically significant, the dots are colored red. Otherwise, they are colored gray. 
   }
\label{fig:NodeHistoryHist}
\end{figure}

We will call the  repayment history features  $X_{FinHist}$, as mentioned in section \ref{sec:featEng}. Figure \ref{fig:NodeHistoryHist}, shows the comparison between the experiments \textbf{E2} and \textbf{E1} to analyze the effect of incorporating the attributes $X_{FinHist}$ to the creditworthiness assessment problem. 

The repayment history features affect creditworthiness assessment performance. Discrimination power, measured as KS, increases as customer history increases and repayment features recollect the borrower's payment behavior. The most meaningful improvements for personal scoring and Business Scoring occur six months after the credit is granted. When performance is measured based on AUC, this relationship is not clear, and the benefits of incorporating repayment history features are observed from the second month. The preceding confirms the importance of incorporating repayment history features.

\subsection{Experiment \textbf{E3}:  borrower credit history, repayment features and social interaction features}

\begin{figure}[ht]
  \centering
  \noindent
  \resizebox{0.85\textwidth}{!}{
  \begin{tabular}{cc}
    \subfigure[Business Scoring (KS)]{\includegraphics[width=0.5\linewidth]{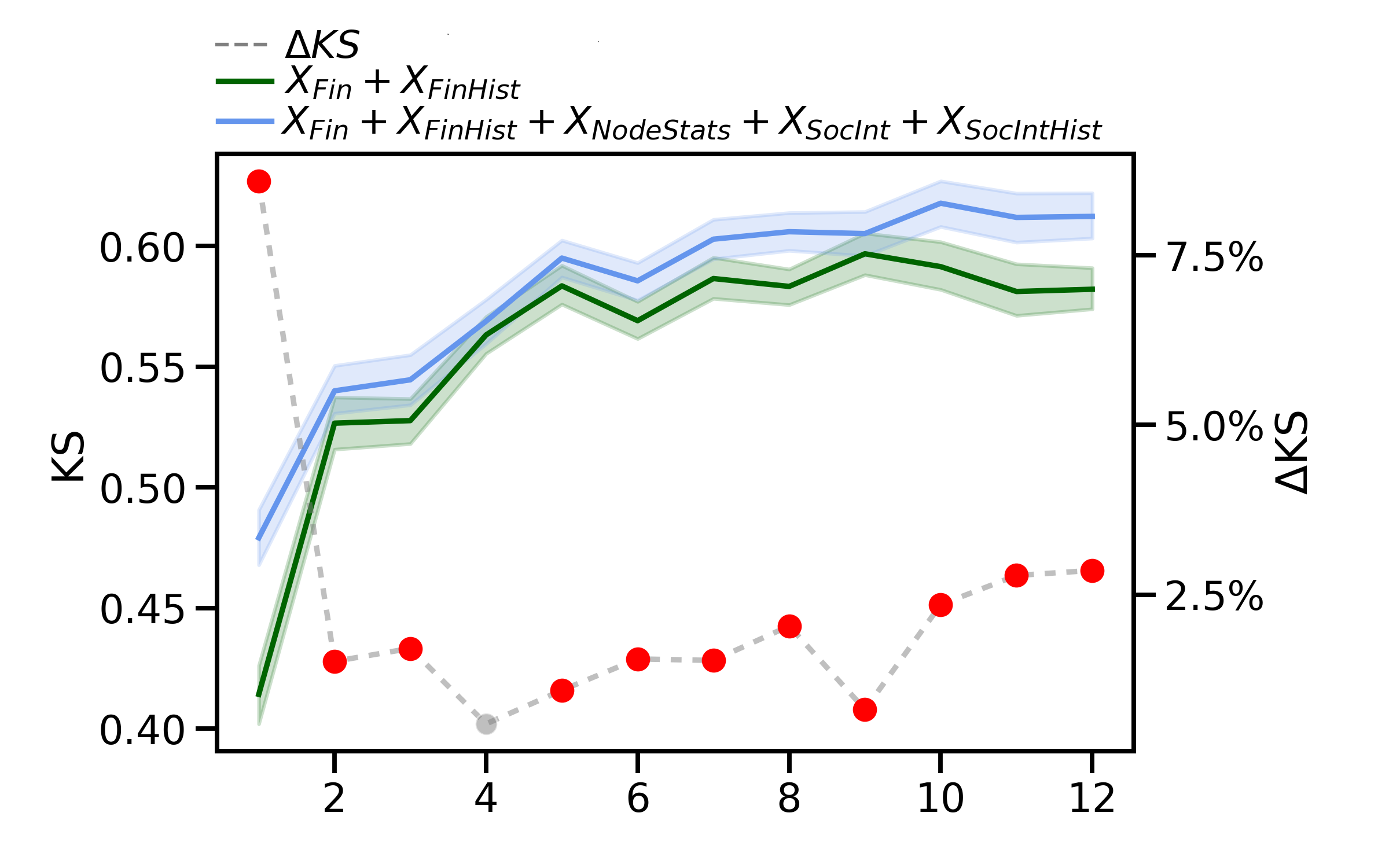}}&
    \subfigure[Business Scoring (AUC)]{\includegraphics[width=0.5\linewidth]{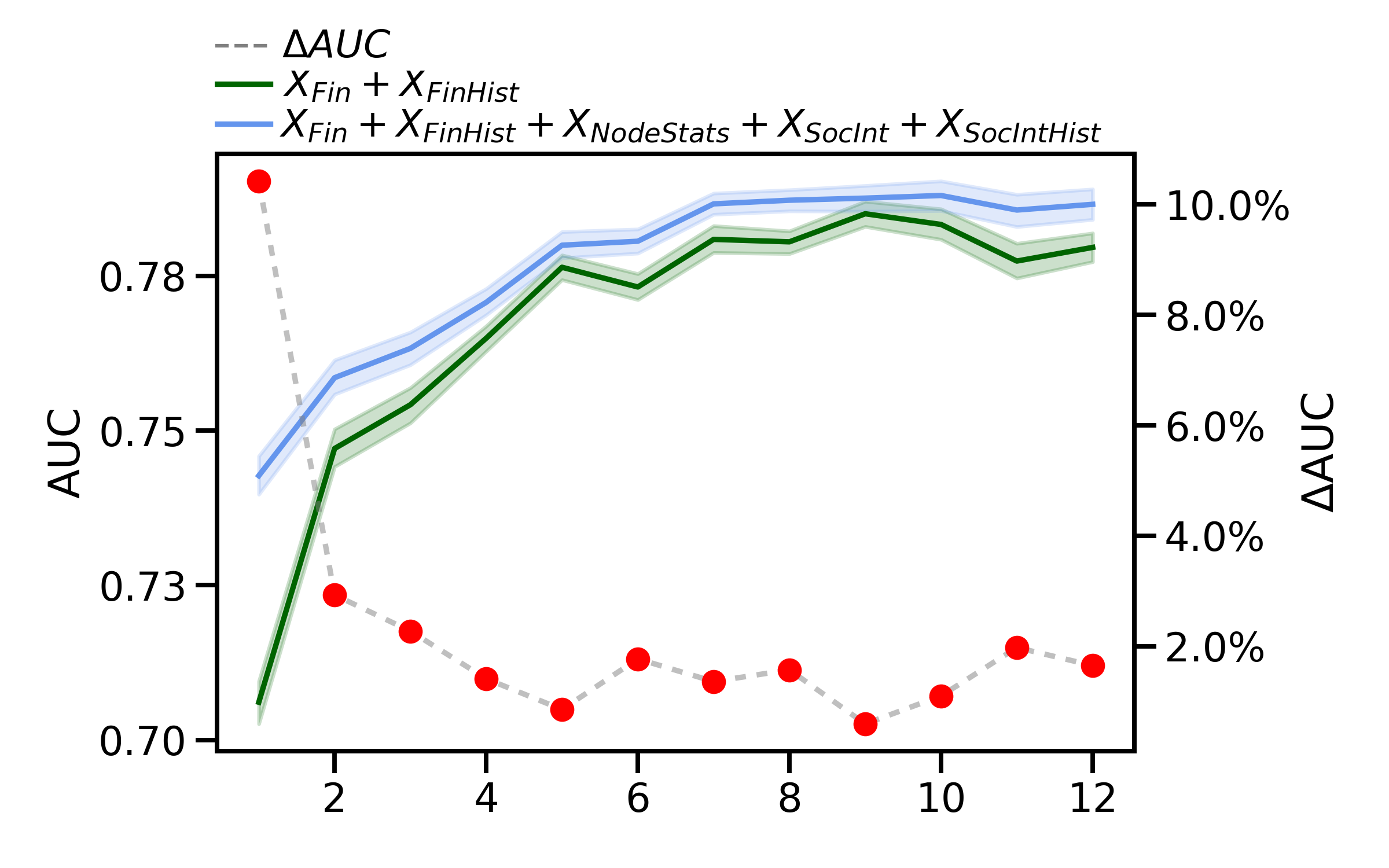}} \\
        \subfigure[Personal Scoring (KS)]{\includegraphics[width=0.5\linewidth]{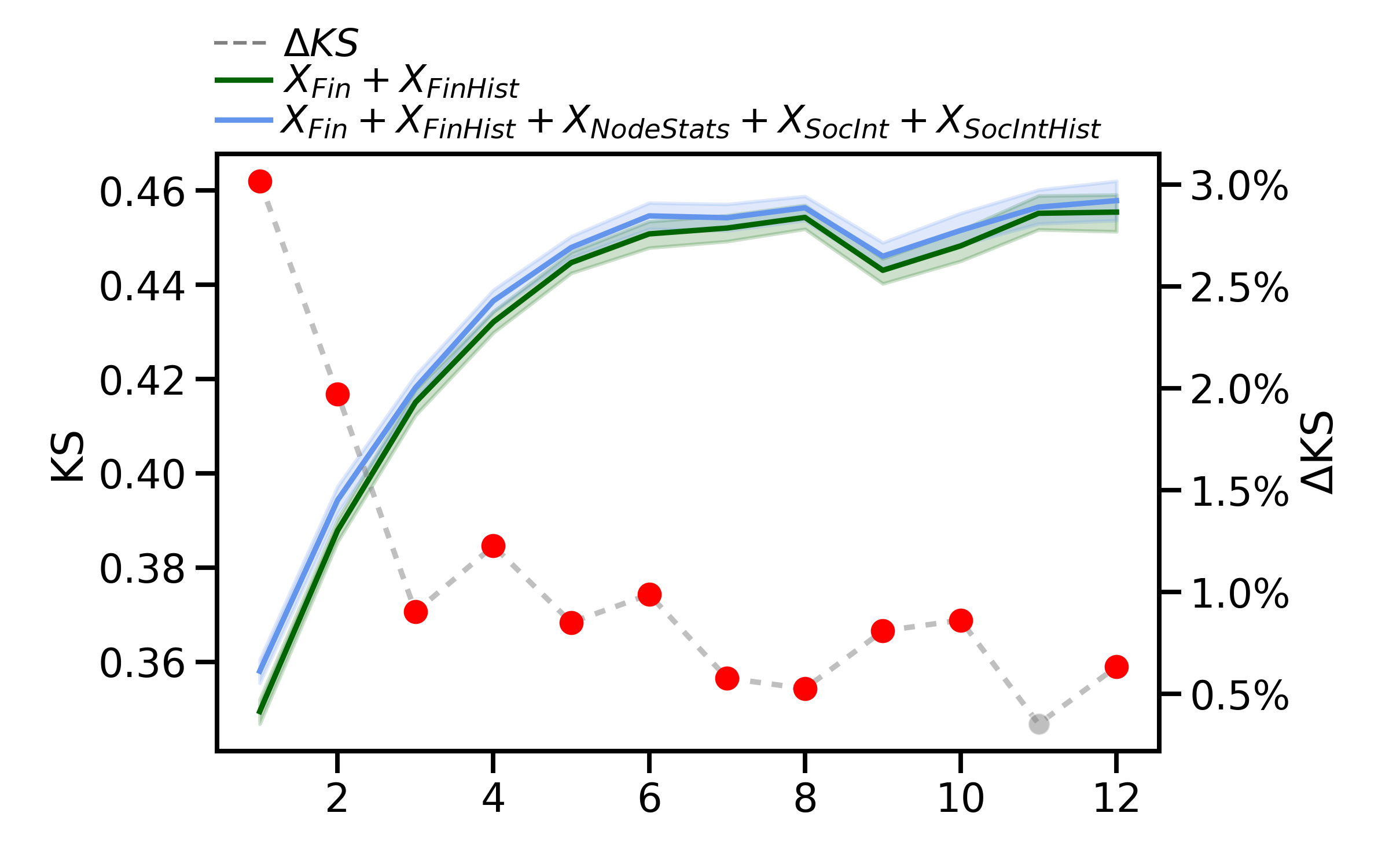}}&
    \subfigure[Personal Scoring (AUC)]{\includegraphics[width=0.5\linewidth]{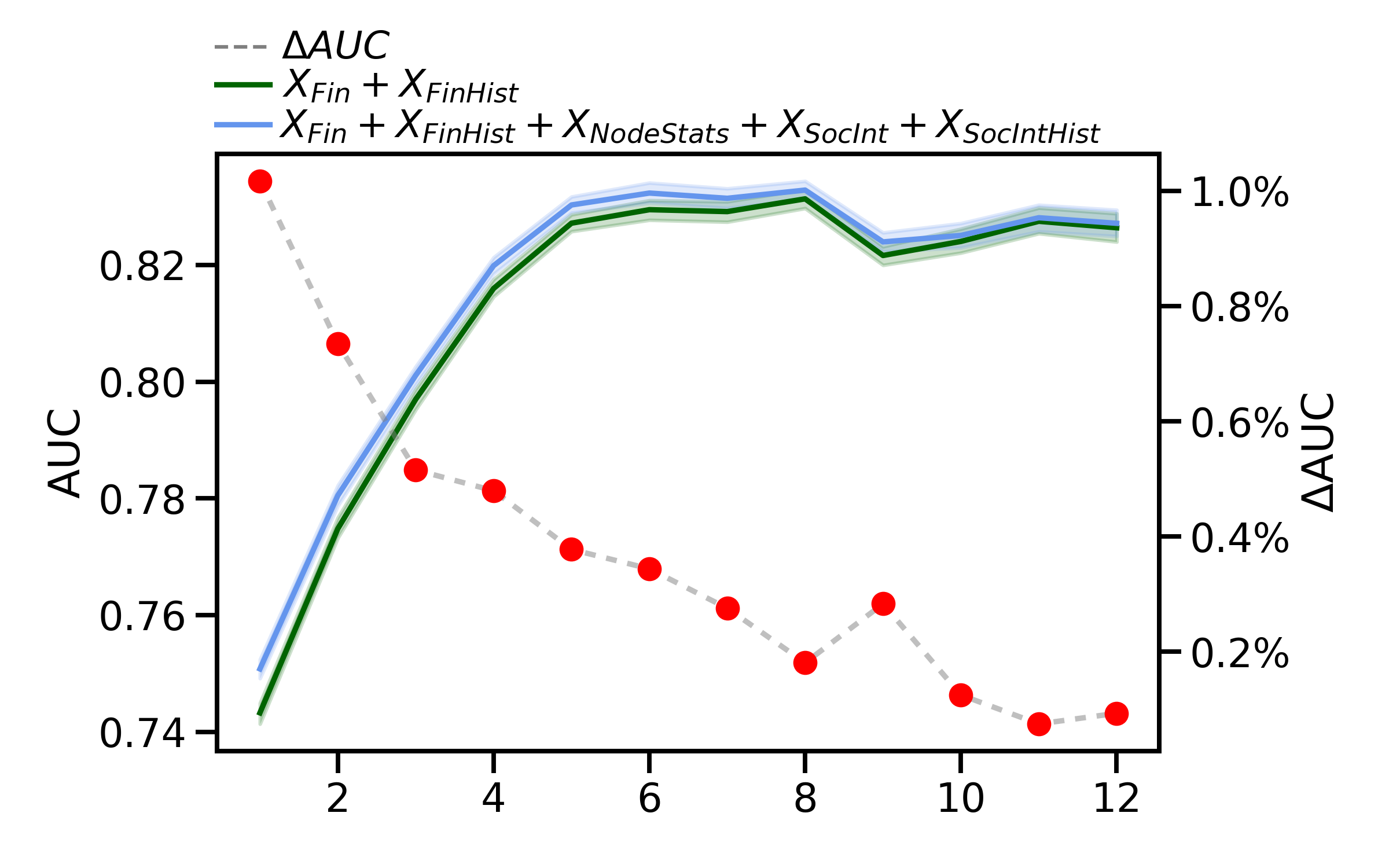}} 
  \end{tabular}}
 \caption{Results in terms of KS and AUC scores for the Business Scoring and Personal Scoring Problem. The X-axis displays the number of months elapsed since the loan granting. The blue line and the green line show the creditworthiness assessment performance (left Y-axis) for Experiment \textbf{E3} and Experiment \textbf{E2}. The dotted gray line (right Y-axis) shows the percentage increment between  \textbf{E3} and \textbf{E2}; when this increment is statistically significant, the dots are colored red. Otherwise, they are colored gray.}
\label{fig:NodeHistoryNetwork}
\end{figure}

One of the questions that motivate this study is to delve into the value delivered by social interaction features, and the impact on creditworthiness assessment performance as the borrower's credit history and repayment features become available. Figure \ref{fig:NodeHistoryNetwork} shows the comparison of experiments \textbf{E3} and \textbf{E2}; this comparison allows us to analyze the added value of social interaction features as the credit history and repayment behavior becomes available.

Incorporating alternative data also allows increasing the creditworthiness assessment performance. The most significant increase is observed in the first month when we face an application scoring problem and the borrower's credit information is not available or it does not exist. 

In personal scoring, social interaction features increase the power of discrimination during the 12 months of studies. However, its impact decreases as the borrower's credit history is available, and repayment features express the borrower's payment behavior better. 
In business scoring, social interaction features increase by about 8\% and 10\% for KS and AUC, respectively. This significant enhancement in discrimination power occurs at loan applications, a clear sign that the creditworthiness assessment of a firm should not only count on its features. This assessment should consider the behavior and attributes of its owners and analyze its supply chain with customers, suppliers, and employees.

\subsection{Importance of social interaction features over time}\label{res:featureImportance}

This section analyzes the borrower's ego network characteristics, i.e, social interaction features and how this impact varies as credit history and repayment behavior become available. To do this, we consider the relative importance of each attribute in order to predict creditworthiness; this relative importance is grouped into two categories, borrower features ($X_{Fin} + X_{FinHist}$) and the features obtained from social network data, i.e, node statistics and social interaction features ($X_{NodeStats} + X_{SocInt} + X_{SocIntHist}$). For each feature, the importance is calculated as the Shapley Values' average from a subset of the dataset and then aggregated according to the two categories previously defined. The Shapley values were obtained using a tree-based SHAP Explainer \citep{Lundberg2017}.

\begin{figure}[ht]
  \centering
  \noindent
  \resizebox{0.85\textwidth}{!}{
  \begin{tabular}{c}
    \subfigure[Business Credit Scoring]{\includegraphics[width=0.95\linewidth]{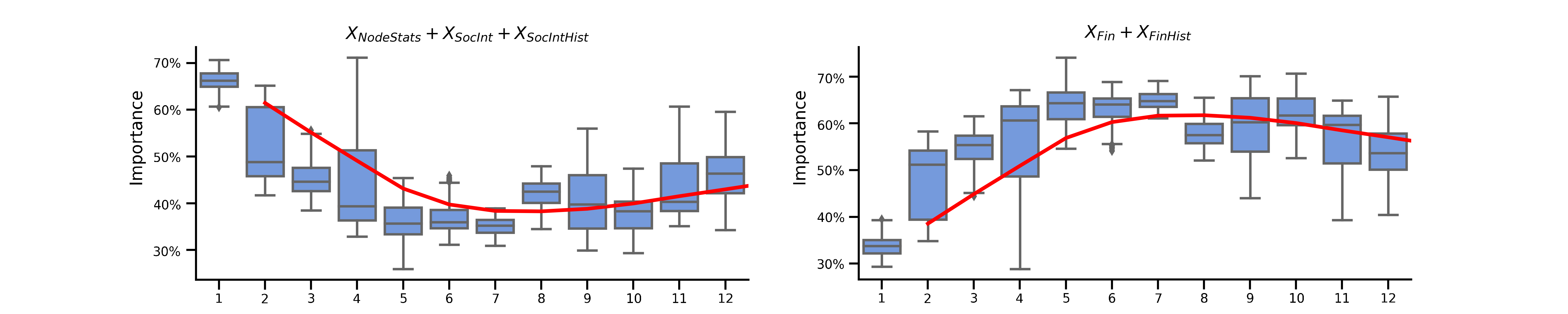}}\\
    \subfigure[Personal Credit Scoring ]{\includegraphics[width=0.95\linewidth]{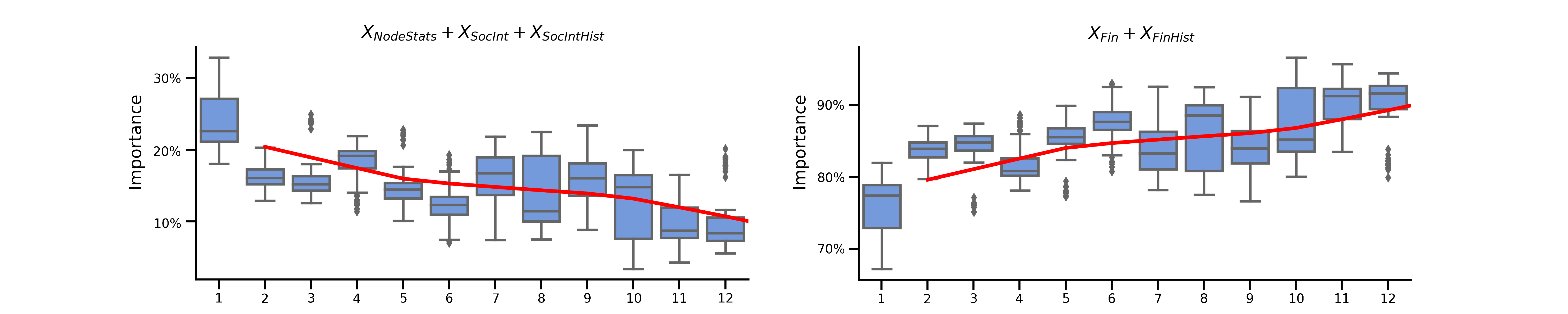}} 
  \end{tabular}}
\caption{Feature Importance Analysis using Shapley Values. Figure (a) presents the Business Scoring problem and Figure (b) Personal Scoring Problem, in both using the Experiment \textbf{E3} feature set. The features are grouped into two categories, the borrower's features ($X_{Fin} + X_{FinHist}$), and the social interaction features ($X_{NodeStats} + X_{SocInt} + X_{SocIntHist}$). The X-axis displays the number of months elapsed since the loan granting. The Y-axis shows the relative feature importance. The boxplots show the feature importance in the 10-fold cross-validation, and the red line is a LOWESS regression fitted using these results.}
\label{fig:ImpShap}
\end{figure}

The Figure \ref{fig:ImpShap}(a) show the feature importance using Shapley values for the business scoring problem and the Figure \ref{fig:ImpShap}(b) for personal scoring.

The importance of network features in business scoring is 63.8\% in the first month, the same month where this information generates its maximum discrimination power enhancement. Something similar happens in personal scoring; however, network information is less important, and so is its increase in discrimination power.

Figure \ref{fig:ImpShap}(a) shows the importance of social interaction features in the creditworthiness assessment; when a company is applying for a loan, this alternative data contributes the most to the credit evaluation. Its value decreases as the company's information become available during the following months; despite this decrease, the importance of social interaction features stabilizes around 40\% from the first six months. This result confirms what is already known by practitioners. The credit evaluation of a firm, especially in small and medium companies, must consider its owners and the business ecosystem where the company interacts.

On the other hand, in personal scoring, the importance of social interaction features diminishes almost linearly as time passes (See Figure \ref{fig:ImpShap}(b)). The most significant enhancement persists at application, but unlike business scoring, the importance of these attributes is considerably smaller, as is their increase in discrimination power. Parental relationships and marriages do not have the same impact on creditworthiness assessment as transactional and economic relationships on business scoring. Despite this, social interaction features allow increasing the power of discrimination. They are fundamental support in the financial inclusion of those people whom traditional credit scoring models cannot evaluate since they do not have a credit history.

\begin{figure}[ht]
  \centering
  \noindent
  \resizebox{1\textwidth}{!}{
{\includegraphics[width=0.7\linewidth]{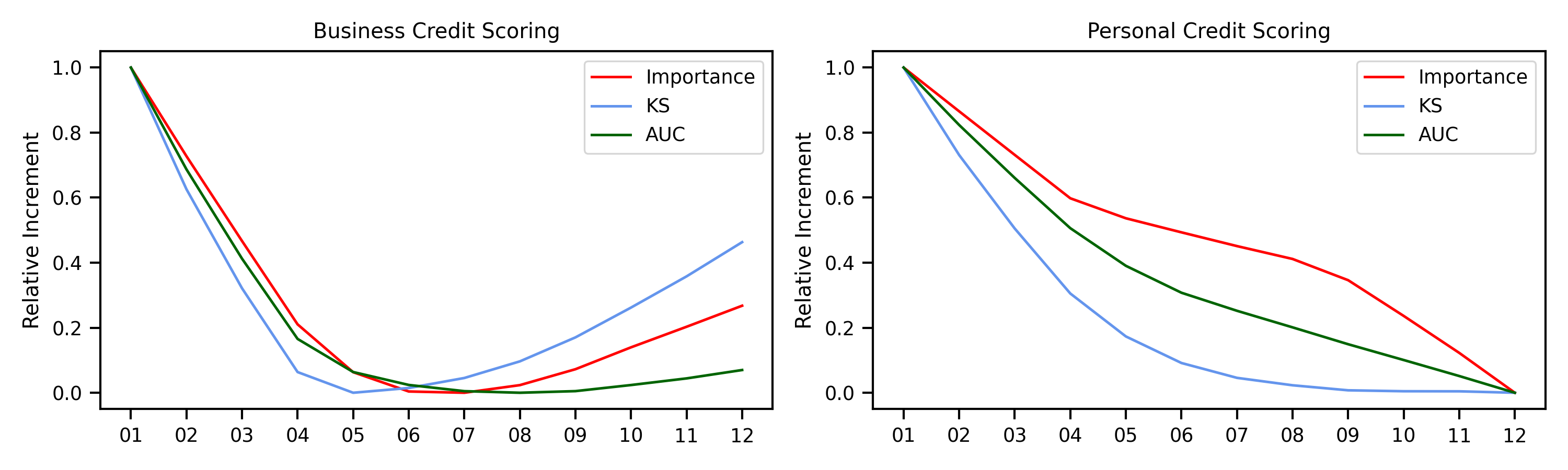}}
}
\caption{Feature importance and predictive power relationship. (Left) Business Scoring problem; (right) Personal Scoring Problem.
in both using the Experiment \textbf{E3} feature set. The blue and green lines show the relative increase between \textbf{E3} and \textbf{E2} experiments for the KS and the AUC. The red line is the Social Interaction features' importance in experiment \textbf{E3}. A \textit{MinMaxScaler} was applied to all series to limit the results between 0 and 1. The X-axis corresponds to the months elapsed after the first loan granting. The Y-axis shows the relative increment.}
\label{fig:ImpTrend}
\end{figure}

An interesting relationship to analyze is the one presented in Figure \ref{fig:ImpTrend}. This figure shows on the same scale the increase in discrimination power measured in KS and AUC and the importance of social interaction features. A high correlation is observed between the increase in discrimination power and the importance of social interaction features in both cases. In business scoring, this correlation is almost perfect during the first six months of the study, and in personal scoring, a strong pseudo-linear correlation is observed. This relationship shows that the contribution of the social interaction features in the creditworthiness assessment is directly translated into an increase in discrimination power.

\section{Conclusions}\label{sec:Conclusions}

This study analyzes how credit history, loan repayment features, and social network data influence the performance of credit scoring models. We use traditional financial data and graph data originating from borrowers' economic and social interactions. Additionally, our novel dataset allows us to analyze all the financial behavior of individuals and companies from the moment they obtain their first loan until 24 months after it. Furthermore, we analyze the performance dynamics based on the results of multiple independent creditworthiness assessment models trained with time-dependent datasets. These models are trained with features representing the borrower's credit history, repayment behavior, and social interactions. The performance of these models is measured in terms of AUC and KS; the feature importance is quantified using Shapley values.

Our findings showed that as more borrowers' credit history is available, creditworthiness assessment performance increases at a decreasing rate. This effect is observed up to six months from the loan granting, when it stabilizes. This finding is meaningful since it reduces the temporal extent of the datasets necessary to train and research behavioral credit scoring models. Also, it increases the population that can be part of these models from the necessary 12 months to only six months. Furthermore, it enhances financial inclusion and leverages second chance banking allowing those borrowers with good credit behavior but with a negative credit history to have a briefer reintegration into the financial system.
An additional noteworthy finding is that the features that summarize the borrower's repayment behavior, the repayment history features $X_{FinHist}$, enhance the creditworthiness assessment performance, especially after the first six months, and consequently, they increase performance when the contribution of the credit history stabilizes.
Finally, The social interaction features allows higher performance and its most significant added value when the borrower is applying for the loan. In personal scoring, this effect decreases almost entirely as the customer's history is available. In business scoring, the increment in discrimination power by incorporating social network features remains stable, at least during the first year.

The results obtained allow us to analyze the dynamics of creditworthiness assessment performance and how it is influenced by the borrower's credit history, the repayment behavior, and the social interaction features. These results are important since it proposes a six-month performance window, reducing it from the current recommendations of 12 months. In addition, they show us how the importance and discrimination power enhancement of social interaction features changes over time. Both insights allow us to improve credit risk management by establishing when and how long to use social network data; similar conclusions are drawn about the performance windows and the contribution of repayment features.

Our study suggests numerous lines of investigation. First, we would like to extend our research period; our 24-month dataset only allows us to have insights from the first 12 months of the borrowers' behavior. Based on our results, it is feasible that the impact of social interaction networks stabilizes after 12 months in the business scoring problem. A more extended observation period would also allow us to study mortgages that have a slower evolution than consumer and commercial credits. Secondly, we would like to understand what happens in other domains, either using other types of networks or studying microcredits or peer-to-peer lending. Finally, in this work, we study the creditworthiness assessment performance dynamics through multiple independent models trained with time-dependent datasets; we would like to design an framework that inherently handles time dependency.

%%%%% Acknowledgments
\section*{Acknowledgments}

This work would not have been accomplished without the financial support of CONICYT-PFCHA / DOCTORADO BECAS CHILE / 2019-21190345. The second author acknowledges the support of the Natural Sciences and Engineering Research Council of Canada (NSERC) [Discovery Grant RGPIN-2020-07114]. This research was undertaken, in part, thanks to funding from the Canada Research Chairs program.
The last author thanks the partial support of  FEDER funds through the MINECO project TIN2017-85827-P and the European Union’s Horizon 2020 research and innovation program under the Marie Sklodowska-Curie grant agreement No 777720.

%%%%%%%%%%%%%%%%%%%%%%%%%%%%%%%%%%%%%%%%%%%%%%%%%%%%%%%%%%%
\bibliographystyle{apalike}

\end{document}